\documentclass{PoS}

\usepackage{amsmath,amssymb,amsfonts}
\usepackage{slashed}%
\usepackage{mathrsfs}

\renewcommand{\epsilon}{\varepsilon}
\renewcommand{\mathcal}{\mathscr}

\newcommand{\hhref}[1]{\href{http://arxiv.org/abs/#1}{arXiv:#1}}
\def\beq{\begin{equation}} 
\def\eeq{\end{equation}}

\title{EWSB Theory on the Eve of Higgs Boson Exclusion/Discovery}

\ShortTitle{EWSB Theory}

\author{{Slava Rychkov}\\
        Laboratoire de Physique Th\'{e}orique, \'{E}cole Normale Sup\'{e}rieure, Paris, France\\
\& Facult\'{e} de Physique, Universit\'{e} Pierre et Marie Curie, Paris, France}

\abstract{A personal feeling of where we stand theoretically in the puzzle of EWSB, faced with the impressive limits on new physics set at the 2011 summer conferences.}

\FullConference{The 2011 Europhysics Conference on High Energy Physics, EPS-HEP 2011,\\
		July 21-27, 2011\\
		Grenoble, Rh\^one-Alpes, France}

\begin{document}

\section{Introduction}

The original title proposed by the organizers was ``New Theories for the Fermi Scale.'' This presupposes a steady influx of interesting new theories, which is not really true. Necessarily, this talk is a mix of new, recent, old, and very old.
I divided it into two parts. I first recount some basic truths and cherished beliefs of the field, and contrast them with some recent `heresies' which try to challenge them. In the second part, I comment on the status of several most important scenarios, paying attention in particular to new limits or the absence thereof. Then I conclude.

\section{Basic Tenets and Heresies}
\subsection{In Naturalness We Trust}
We start with Naturalness; how could it be otherwise. Much of the theoretical thought in the last 30 years has been moved by a conviction that fundamental scalars are unnatural. The simplest case is the $\lambda\phi^4$ theory:
\beq
\mathscr{L}=(\partial \phi)^2+m^2\phi^2+\lambda\phi^4\,.
\eeq
As often said, making this theory valid up to the energies $\Lambda\gg m$ requires finetuning.

It is less often mentioned that this fact has actually been experimentally verified, in condensed matter physics.
Ferromagnets near the Curie point are described by the same Lagrangian, in three dimensions. This is the Landau-Ginzburg theory of phase transitions. In condensed matter language, the inverse cutoff $\Lambda^{-1}$ corresponds to the atomic spacing $a$, while the inverse mass $m^{-1}$ is the spin correlation length $\xi$:
\beq
\Lambda^{-1} \longleftrightarrow a,\qquad m^{-1} \longleftrightarrow \xi\,.
\eeq
So, the limit $m\ll\Lambda$ corresponds to $\xi\gg a$, which is the condition for being near the critical point.

\begin{figure}[htbp]
\begin{center}
\raisebox{-2em}{\includegraphics[scale=0.35]{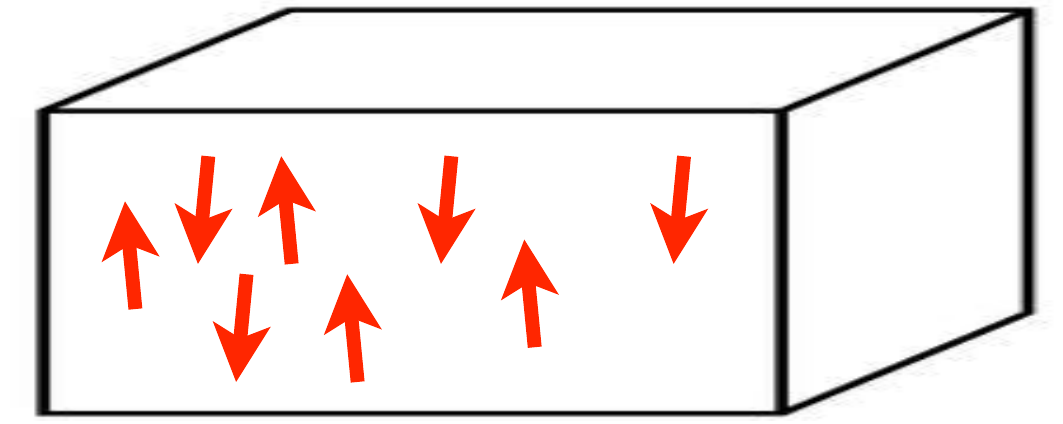}}\hspace{1cm}$\Longrightarrow$\hspace{1cm}
\raisebox{-3.5em}{\includegraphics[scale=0.35]{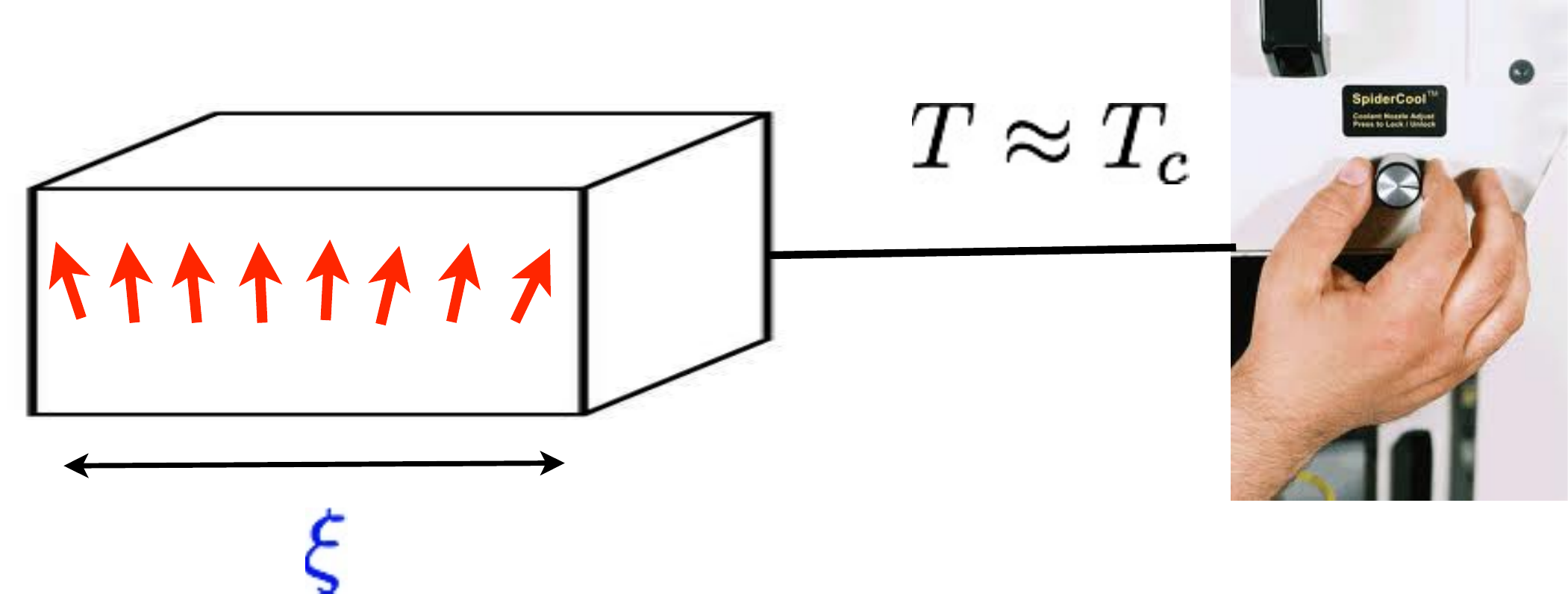}}
\caption{Finetuning in condensed matter physics. }
\label{ferro}
\end{center}
\end{figure}

Now, it is a fact that a random piece of ferromagnet is unlikely to be found near the critical temperature. The limit $T\to T_c$ requires the presence of an experimenter, who finetunes the temperature turning a knob (Fig.\ \ref{ferro}).
This is precisely the finetuning we are talking about. 

If we believe, as most of us do since the work of Ken Wilson and Alexander Polyakov, that particle physics is, fundamentally, condensed matter physics of the vacuum, we come to the conclusion that something has to be done about the hierarchy problem of the Standard Model. That is, barring the existence of an Experimenter.  

\subsection{Exit strategies if Naturalness fails}
Some people remain skeptical about Naturalness. Maybe they do not like the condensed matter analogy, or maybe they just want to put a stake in a virgin field, which is a reasonable thing to do. As a result there is already a number of exit strategies in the literature, in case Naturalness fails and the LHC sees the Higgs boson and nothing else.

In principle, I have nothing against these attempts as long as they lead to observable predictions.\footnote{On the other hand, I refuse to be drawn into a vacuous discussion of the finetuning argument being wrong because you can renormalize the SM without ever encountering quadratic divergences, like in dimensional regularizations, etc.} There exist a few rather neat models driven by minimality. For example, Ref.~\cite{nuMSM} proposes to extend the SM by three right-handed neutrinos ($\nu$MSM), with masses in keV-MeV range. In such economical setup, one manages to solve all the problems of particle physics: neutrino oscillations, Dark Matter abundance, and baryogenesis.

Another interesting example is the Minimal Dark Matter model \cite{MDM}. By adding to the SM, assumed to be valid up to a very high scale, just one fermionic fiveplet of $SU(2)$, one gets a Dark Matter candidate which is stable basically for the same reason as the proton: there is no renormalizable operator which can lead to its decay. This is to be contrasted with a $\mathbb{Z}_2$ symmetry normally invoked to ensure stability of most other DM candidates.

Then there exists an escape route of the second kind, which looks more like a suicide when you first hear about it.
Here you do take the existence of the hierarchy problem for real, but assume that it is solved by environmental selection in a Landscape of vacua. If you assume that the distributions on the Landscape are peaked here or there, you can make predictions. The person who probably thought most concretely about this in recent years from the particle physics perspective is Lawrence Hall with collaborators. E.g., in 2006 they predicted \cite{Hall1}
\beq
m_H=115\pm 6\ \text{GeV}\,, 
\eeq 
while in 2009 they came up with an even more precise prediction \cite{Hall2}
\beq
m_H=141\pm 2\ \text{GeV}\,.
\eeq 
Obviously, different assumptions lead to different predictions. What worries people most about this approach is that if the Higgs mass is the only particle physics measurement left to perform, it may be very difficult to gain confidence in the underlying assumption. 

\subsection{Need for unitarization and Iconoclasm}
I now come to the second basic truth. We all know that the Higgsless SM is incomplete and requires UV cutoff at $\Lambda_{UV}\sim 4\pi v\sim 2\div3$ TeV. Just look at the longitudinal WW scattering. Without Higgs, the tree-level amplitude for $W_LW_L\to W_LW_L$ scattering grows like $s/v^2$ and exceeds unitarity at $\Lambda_{UV}$, signaling strong coupling. Put a not-to-heavy Higgs boson back, and the amplitude, having peaked at $s\sim m_H^2$, drops to a constant value less than one.

This is what happens in the SM, but we believe that this picture is more general. In any theory, WW scattering will first grow linearly, then pass through a resonance region (Higgses, composite vectors of Technicolor, etc.), and then enter a new regime governed by a new, better theory which we would like to discover. At these asymptotically high energies the amplitude will be unitary of course, but to compute it we need to know what this new theory is. 

Sounds reasonable, right? Not so fast; there are people for whom nothing is sacred. It was claimed last year \cite{Dvali} that the Higgsless SM may be UV complete by itself, in a novel sense. No need to change the description, no need to add degrees of freedom. All you need to compute the high energy WW scattering is to solve classical field equations using the Higgsless SM Lagrangian, perhaps supplemented by a higher-derivative term.

There are obvious complaints to be filed about this proposal. First of all, we would like to know what happens in the resonance region, which is the one most relevant for the LHC. Is this also somehow secretly contained in the Higgsless Lagrangian? Second, even at asymptotically high energies, so far there is no concrete computation of, say, WW scattering at 10 TeV.

It is fair to say that the majority of theorists remain deeply skeptical about this proposal. So the experimentalists probably should not worry about it (not yet anyway). But remember, the organizers wanted me to talk about new theories. And this is pretty much the only radical idea proposed last year.

\section{Comments on Models}

\subsection{Supersymmetry}

\begin{figure}
\raisebox{0em}{\includegraphics[scale=0.7]{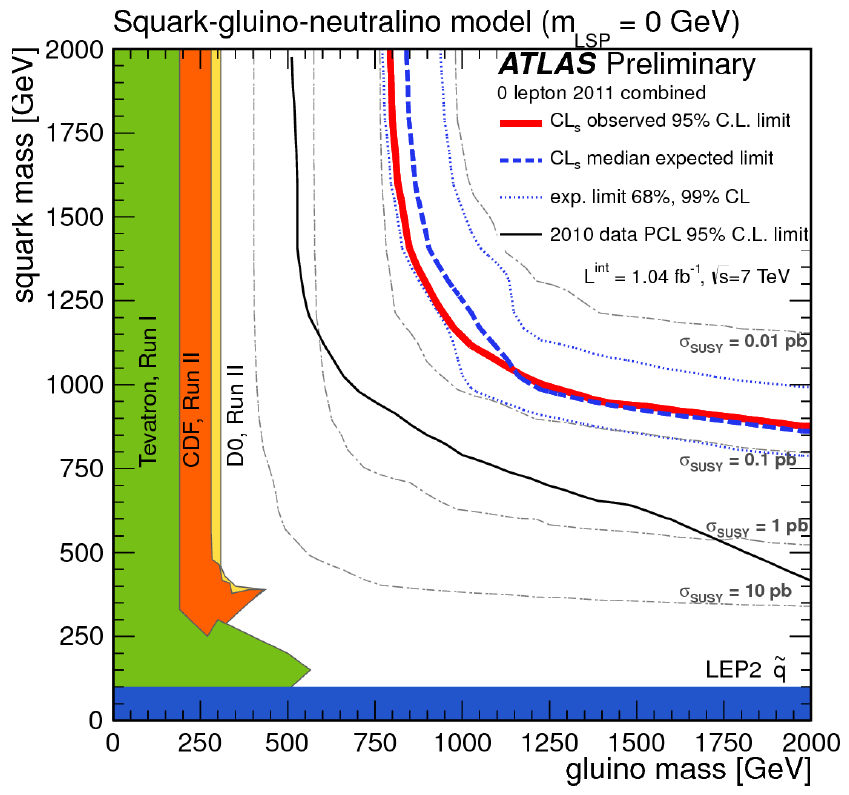}}\qquad
\raisebox{0em}{\includegraphics[scale=0.7]{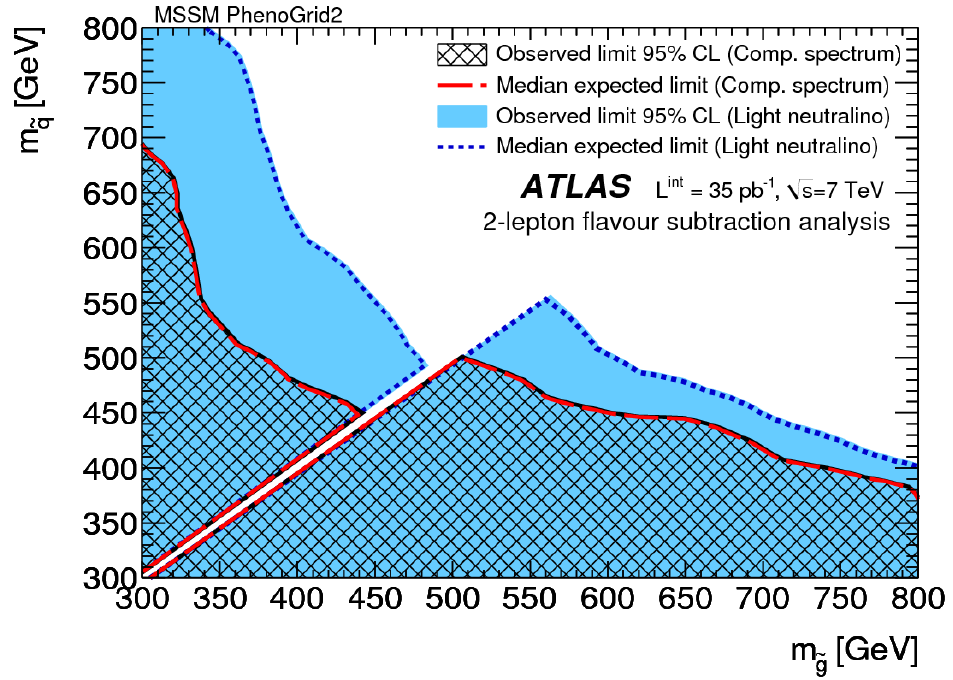}}
\caption{Limits on simplified SUSY models corresponding to the decay topologies corresponding to the dominant decays $\tilde g\to q\tilde q_{1,2}$, $\tilde q_{1,2}\to q+\text{LSP}$ giving light quark jets plus $\slash\!\!\!\!E_T$ in the final state (left) and longer decay chains $\tilde q_{1,2}\to q\tilde\chi_2^0$, $\tilde\chi_2^0\to l\tilde l$,
 $\tilde l\to l+\text{LSP}$, giving also leptons (right). See talks by Taffard and Vivarelli. The first two generations of squarks are assumed to have equal mass in these analyses.}
\label{susy-limits}
\end{figure}

We have seen many new SUSY limits presented at this conference, see Fig.~\ref{susy-limits}. It's becoming customary for the collaborations to present results in terms of physical parameters of simplified models \cite{Simplified} corresponding to the relevant topologies. This is both more general and more immediately useful than exclusion plots in mSUGRA/CMSSM parameter space.\footnote{In the same vein, kudoz to ATLAS for presenting limits on dijet resonances as a function of the resonance width rather then tailoring to the model zoo, see Charlton's talk.} I hope that as time goes by we will see fewer and fewer mSUGRA plots.

In any case, we are faced with impressive bounds going into a TeV range. What do we learn from them? The answer is as follows (see Papucci's talk). On the one hand, plain vanilla SUSY models, like MSSM with flavor-universal scalar soft masses, so loved by people doing global fits, are being pushed further into a corner. One could venture as far as to say that CMSSM is basically out. On the other hand, there remain a few theoretically motivated scenarios which are still very poorly constrained by the existing searches. These are scenarios which do not give rise to lots of missing energy, e.g.~R-parity violation but not only it, as well as scenarios in which first and second generation squarks are much heavier than the third generation ones. 

\begin{figure}[htbp]
\begin{center}
\raisebox{0em}{\includegraphics[scale=0.5]{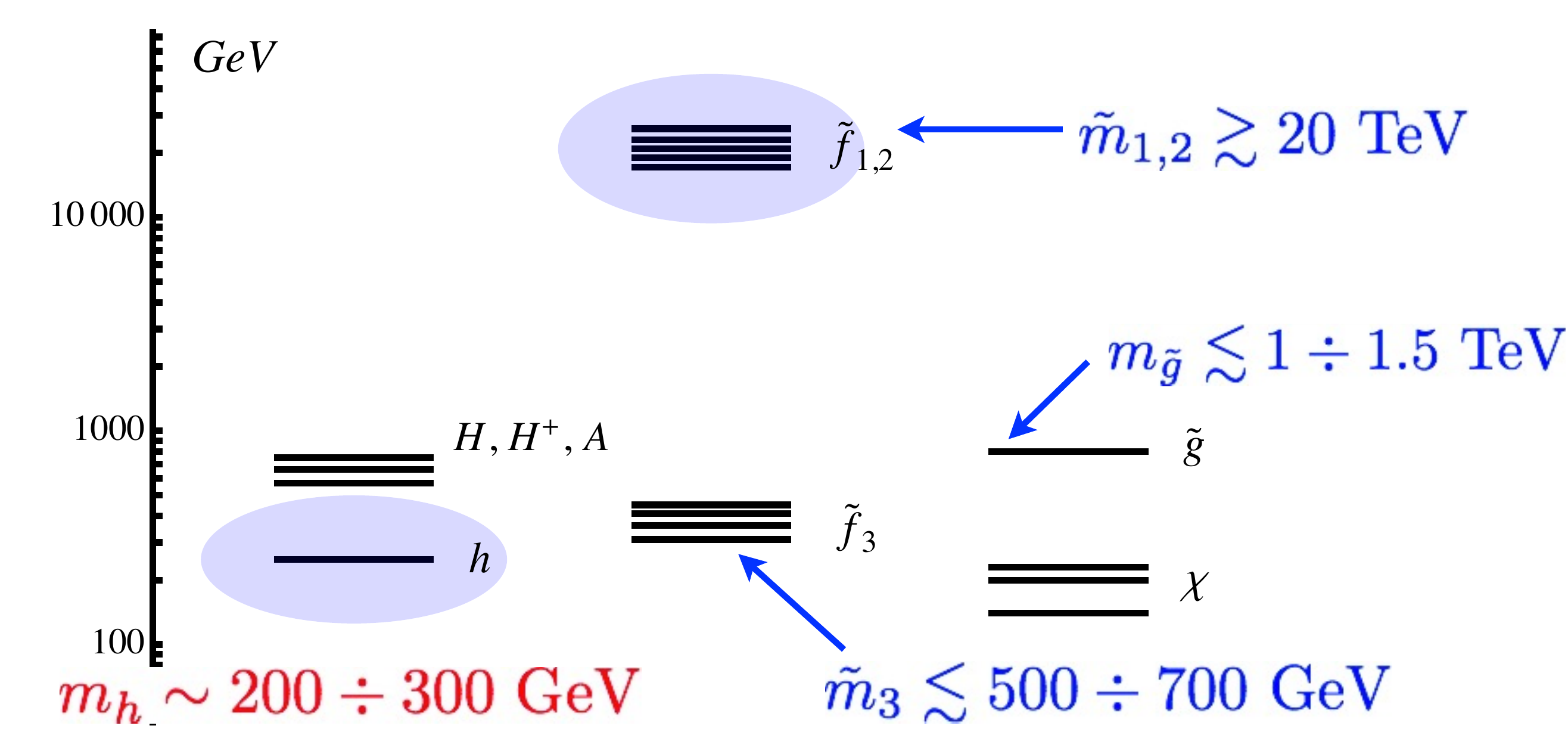}}
\caption{A motivated non-standard SUSY spectrum with first two squark generation decoupled.}
\label{non-standard}
\end{center}
\end{figure}

The latter idea, dubbed Effective SUSY in the 90's \cite{Cohen} makes a lot of sense from the point of view of naturalness (which only demands that the stops, having significant coupling to the Higgs, should be ``light"). The Pisa group has done a lot of work on these models in recent years \cite{Pisa}. A typical natural spectrum (Fig.~\ref{non-standard}) has stops at $500\div700$ GeV, gluino at $1\div1.5$ TeV, while first and second generation squarks should be $\gtrsim 20$ TeV for suppressing flavor violation 
effects\footnote{Most recently, they invoked the $U(2)^3$ flavor symmetry. In this case the first two generations  of squarks do not have to be that heavy. Only the flavor-blind phases set a limit via the EDM, and around 3 TeV is enough for maximal phases. I am grateful to R.~Barbieri for emphasizing this to me. With this new twist, one does not have to raise the Higgs mass to maintain naturalness. One just needs to exceed the LEP $114.4$ GeV bound, which may be done with moderate $\lambda$'s.} (see also Straub's talk). It turns out that such heavy $\tilde q_{1,2}$ start feeding significantly into the Higgs mass running at the two-loop level. Maintaining naturalness then requires raising the lightest Higgs boson mass to $200\div 300$ GeV, which can be accomplished by the ``Fat Higgs"/$\lambda$SUSY mechanism \cite{lambda}.
Discovery of sparticles in this model  can be done via the gluino cascades into third generation quarks plus missing energy, which do not seem to be significantly constrained at the time of this conference (see Papucci's talk).

 There's not much else to say. We should wait and see how dedicated searches in the as yet unconstrained scenarios proceed. It is too early to write SUSY off.
  
\subsection{Strong EWSB\protect\footnote{Dedicated to CERN Council Members who, I hear, start being worried why the Higgs boson is not yet discovered.}}
  
\begin{figure}[htbp]
\begin{center}
\raisebox{0em}{\includegraphics[scale=0.3]{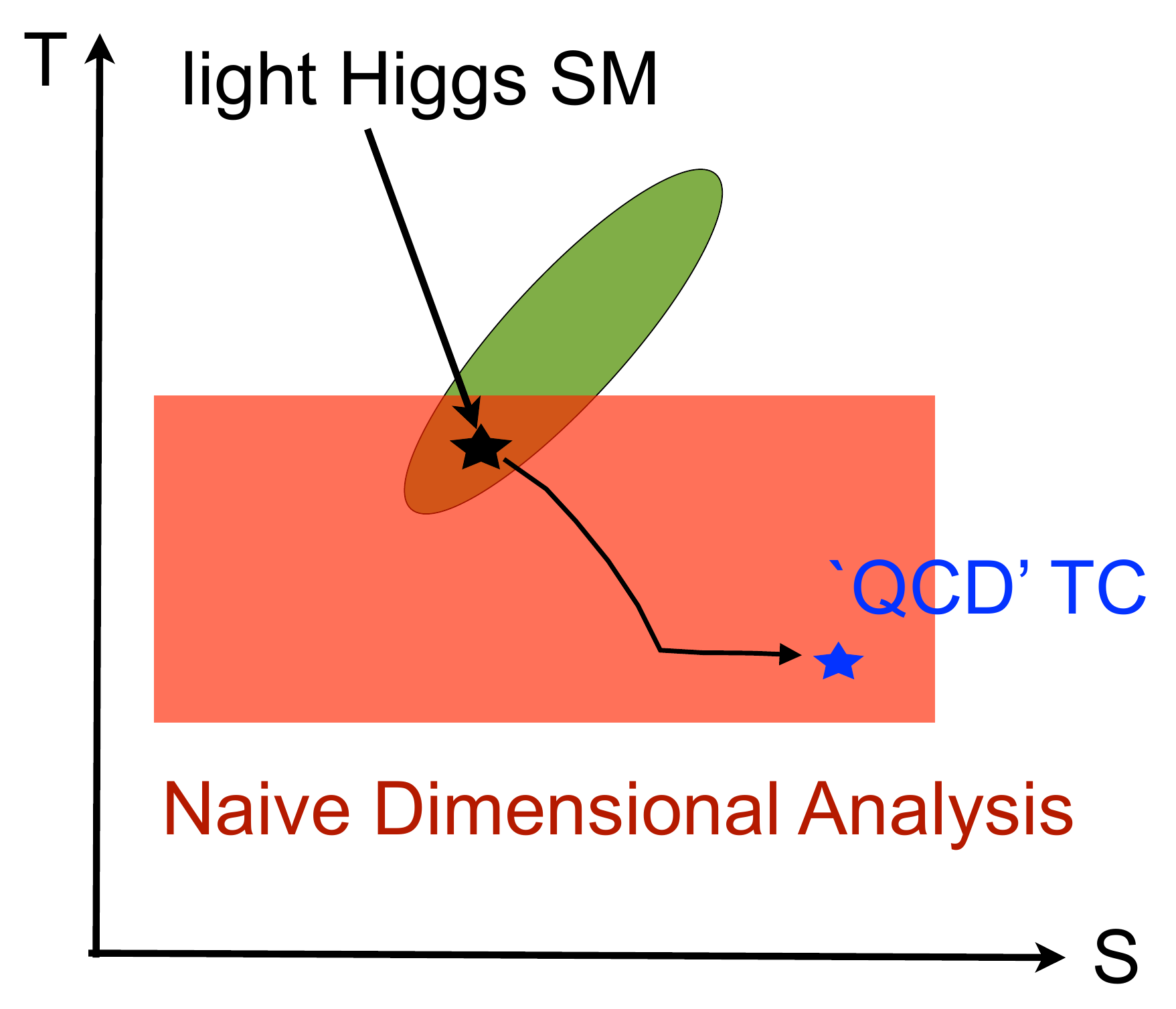}}
\caption{Electroweak precision tests and Technicolor. Caricature adapted from Markus Luty.}
\label{TC}
\end{center}
\end{figure}

I mean of course Technicolor-like theories and their cousins. After LEP these theories fell out of fashion because they are thought to predict large deviations in electroweak precision parameters, especially if TC is QCD-like (Fig.~\ref{TC}). But TC does not have to resemble QCD, in which case we can only do an order-of-magnitude estimate of $S$ and $T$. A no-go theorem has never been proved, and taking into account how little we know about strongly coupled theories, we cannot exclude a 10\% accident which will make them consistent. 

When one mentions TC, people sometimes complain: ``Oh no, it's going to be like QCD all over again. Please, let's have something new.'' However, it's pretty clear by now that if TC is realized, it's UV structure will be quite unlike QCD. And not just because of the electroweak precision tests. Flavor constraints cry out about it.

Recall that the Higgs field of TC theories is a composite operator, and so the Yukawa couplings are no longer dimensionless:
\beq
\frac{y}{\Lambda^{\text{dim}H-1}} H_{TC}(\bar q q)_{SM} 
\eeq
If TC is QCD-like, then $H_{TC}\sim\bar\psi\psi$ a technifermion bilinear, and $\text{dim}H\approx 3\gg 1$. In this case the Yukawas are strongly suppressed and the UV scale $\Lambda$ cannot be very large if, say, we want to reproduce the top, or even the bottom mass. This in turn leads to unacceptably large FCNC.

The way to overcome this well-known problem is to assume that above the EW scale, TC does not transition immediately to a weakly coupled regime, but instead gets stuck near a scale-invariant, strongly interacting fixed point. This kind of behavior was dubbed walking \cite{Holdom}. In such theories, Higgs field dimension can interpolate between 3 and 1.

Actually, much of the older literature has focused on the case $\text{dim}H\approx 2$, when the FCNC problem is mitigated but not really solved. The most economical scenario, allowing to give masses to all SM fermions, including the top, without running into flavor problems requires something like $ \text{dim}H\leqslant 1.5$ \cite{Luty}

\begin{figure}[htbp]
\begin{center}
\raisebox{0em}{\includegraphics[scale=1]{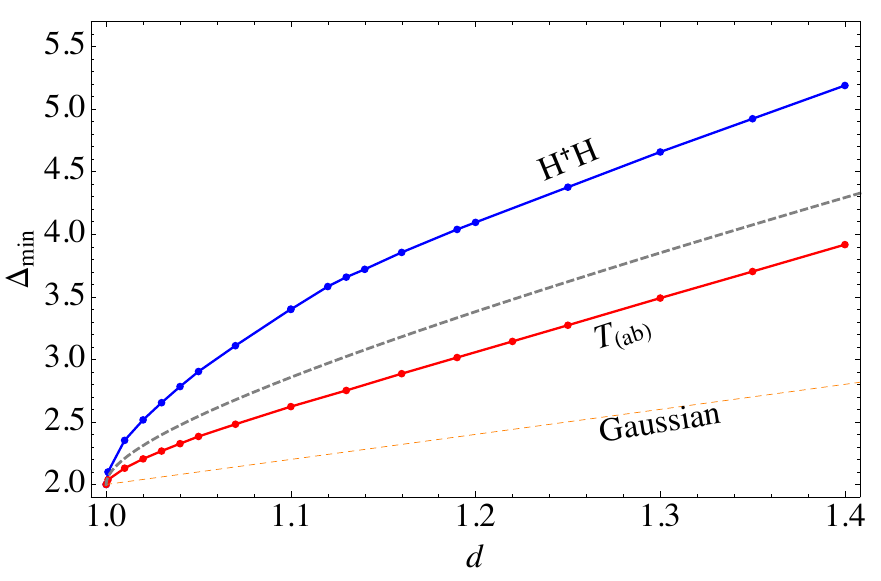}}
\caption{Horizontal axis: the dimension of $H$. Blue curve: the minimal allowed dimension of the singlet scalar operator ($H^\dagger H$). It is assumed that the CFT has an $SU(2)\times SU(2)$ global symmetry. The interesting region is where $\text{dim}H\lesssim 1.5$ while $\text{dim}H^\dagger H\gtrsim 4$. See \cite{Vichi} for the bound and \cite{R} for the theory behind. The bound shown in the conference slides \cite{PSV} is stronger but also does not exclude the Conformal Technicolor scenario.}
\label{Vichi}
\end{center}
\end{figure}

One may wonder if such a large drop in dimension (from 3 to 1.5) can be possible at all, consistently with the IR stability of the fixed point. Personally, I find it interesting that rigorous bounds on dimensions of operators in Conformal Field Theories constrain but do allow this scenario (Fig.~\ref{Vichi}).

Expected signals of strong EWSB are resonances at a few TeV. For example, one may expect vector resonances (sometimes called techni-$\rho$) with mass $M_\rho\sim 1\div 3 TeV$. Such vectors may basically replace the Higgs boson in moderating the growth of WW scattering amplitude (partial unitarization). It is interesting that they can accomplish this task while remaining an order of magnitude more narrow than a would-be Higgs boson of the same mass \cite{Bagger,us}.

As is well known, techni-$\rho$'s decay to WW,WZ and are produced in WW fusion and in DY. However, since they couple to light SM fermions not directly but rather only due to a small mixing with the SM gauge bosons, the coupling to quarks is expected to be strongly suppressed, and even new limits from sequential $Z'$ and $W'$ searches reported at this conference do not apply. These signals may require O(100 fb$^{-1}$) luminosities.

In the scalar sector, Higgs-boson-like isospin singlets are expected to be very wide due to strong decays into WW,
and as elusive as the $\sigma$-boson in QCD. However, there can be also isotriplets or negative parity isosinglets, with a significant branching ratio into three gauge bosons. In conformal technicolor scenario, they can be produced efficiently in gluon fusion via the top loop, and provide a promising experimental signature \cite{EvansLuty}.

If a light Higgs boson is discovered, the TC story will be relegated to history books. This cannot be said about the next and last topic of my review.

\subsection{Composite Higgs boson}

More precisely, I mean scenarios with a Higgs boson as a pseudo-Nambu-Goldstone boson (PNGB) of a spontaneously broken global symmetry. This topic has been discussed in plenary talks by Giudice \cite{Giudice} at EPS 2009 and Grojean\cite{Grojean} at EPS 2011 (see also Rattazzi \cite{Rattazzi} at EPS 2005). My purpose here is to reiterate the strong appeal and promise that theorists see in this idea.

In these models, the onset of truly strong interactions is postponed to a scale a few times higher than the Higgsless strong coupling scale $4\pi v$ (``strongish EWSB"). This allows to make predictions about the dynamics of these models in the LHC energy range, with a relatively small number of free parameters. 

By assumption, this strongish dynamics leaves one or more Higgs bosons as low-energy remnants, naturally light because PNGBs. Derivative interactions of these Higgses are largely controlled by the symmetry. Higgs potential, on the other hand, is controlled by small symmetry breaking terms. One (or perhaps the only) source of such breaking arise from couplings between the strong sector and the rest of the SM (gauge bosons and fermions). 

We don't know what the global symmetry of these models is (experiment will tell). Theoretically, there is a discrete list of possibilities (Table~\ref{cosets}). Notice that all the listed cosets contain at least one $(2,2)$ of the custodial $SU(2)\times SU(2)$. This is the composite Higgs doublet, which must be present among the goldstones in order to realize the breaking of electroweak symmetry via the vacuum disalignment mechanism.
\begin{table}[h!]
	\begin{center}
\begin{tabular}{ccccccc}
\hline
&$G$ & $H$ & $N_G$ & NGBs $\textrm{rep.}[H] = \textrm{rep.}[\textrm{SU}(2) \times \textrm{SU}(2)]$ \\

$\star$&SO(5) & SO(4) & 4 & $\mathbf{4} = (\mathbf{2},\mathbf{2})$ \\

$\star\star$&SO(6) & SO(5) & 5 & $\mathbf{5} = (\mathbf{1},\mathbf{1}) + (\mathbf{2},\mathbf{2})$ \\

$\star\star\star$&SO(6) & SO(4) $\times$ SO(2) & 8 & $\mathbf{4_{+2}} + \mathbf{\bar{4}_{-2}} = 2 \times (\mathbf{2},\mathbf{2})$ \\


&SO(7) & SO(6) & 6 & $\mathbf{6} = 2 \times (\mathbf{1},\mathbf{1}) + (\mathbf{2},\mathbf{2})$ \\

&SO(7) & $\textrm{G}_2$ & 7 & $\mathbf{7} = (\mathbf{1},\mathbf{3})+(\mathbf{2},\mathbf{2})$ \\

&SO(7) & SO(5) $\times$ SO(2) & 10 & $\mathbf{10_0} = (\mathbf{3},\mathbf{1})+(\mathbf{1},\mathbf{3})+(\mathbf{2},\mathbf{2})$ \\

&SO(7) & $[\textrm{SO}(3)]^3$ & 12 & $(\mathbf{2},\mathbf{2},\mathbf{3}) = 3 \times (\mathbf{2},\mathbf{2})$ \\

$\star\star\star$&Sp(6) & Sp(4) $\times$ SU(2) & 8 & $(\mathbf{4},\mathbf{2}) = 2 \times (\mathbf{2},\mathbf{2}), (\mathbf{2},\mathbf{2}) + 2 \times (\mathbf{2},\mathbf{1})$ \\

&SU(5) & SU(4) $\times$ U(1) & 8 & $\mathbf{4}_{-5} + \mathbf{\bar{4}_{+5}} = 2 \times (\mathbf{2},\mathbf{2})$ \\

&SU(5) & SO(5) & 14 & $\mathbf{14} = (\mathbf{3},\mathbf{3}) + (\mathbf{2},\mathbf{2}) + (\mathbf{1},\mathbf{1})$ \\


\hline
\end{tabular} \\
	\caption{Cosets $G/H$ from simple Lie groups, 
	with $H$ maximal subgroup of $G$. For each coset, 
	its dimension $N_G$ and the NGBs representation under $H$ are given \cite{THDM}.}
	\label{cosets}
	\end{center}
\end{table}

Some of the models listed in the table have already been studied. The Minimal Composite Higgs model ($\star$) \cite{ACP,SILH} is the truly minimal realization of this scenarios. The Next-To-Minimal model ($\star\star$) \cite{NMCHM} contains an extra singlet goldstone, which leads to richer phenomenology (see below). Finally, composite realizations of Two-Higgs Doublet Models have also been investigated ($\star\star\star$) \cite{THDM}. We are waiting for the Higgs boson discovery before we proceed further down the list.

Generic predictions for the Higgs boson properties in these models are as follows. The Higgs boson is SM-like and is typically light, since it's mass is zero at tree level and is due only to loop effects. $M_H\lesssim 200$ GeV is usually quoted as the preferred range, but the values as high as 300 GeV cannot be excluded. The Higgs boson couplings to gauge bosons and fermions deviate from their values in the SM by $O(10\div 20)\%$. These changes are correlated: in the minimal model they are controlled by just two parameters, which goes up to three if the top quark has a significant composite component (see below). Interestingly, the sign of these deviations can also be predicted on theoretical grounds. In most cases, it is negative (suppression) \cite{LRV}. Finally, additional goldstones present in the non-minimal model can appear as extra decay channels of the SM-like Higgs boson. For example, the $H\to\eta\eta$ decays, where $\eta$ is the singlet goldstone of the Next-to-Minimal, $SO(6)/SO(5)$ model \cite{NMCHM}. If LHC sees some hints of this kind of deviations from the SM Higgs behavior, we will likely need ILC to fully explore it.

Composite Higgs models have an interesting connection to flavor physics (see Weiler's talk). Typically, they use the so-called ``partial compositeness'' mechanism to give masses to SM fermions \cite{Kaplan}. The mechanism assumes that there are bilinear couplings between SM fermions $\psi$ and vector-like composite fermion resonances $\Psi$ of the form:
\beq 
y_L \bar\psi_L \Psi^{(1)}_R+ y_R \bar\psi_R \Psi^{(2)}_L\,.
\eeq
 Notice that the left and right-handed SM fermions can couple to the right- and left-handed components of, in general, different composite fermions.
These couplings break the global symmetry of the strong sector since SM fermions do not come in complete multiplets of it. As a consequence, contributions to the Higgs boson potential are generated, as mentioned above.

To summarize, the SM fermions pick up their mass as a result of their mixing with heavy composite fermions, which explains the name of this mechanism. Their mass is proportional to the product of left and right mixing angles. The most composite SM fermion is thus the top, whose right-handed component can even conceivably be fully composite.

The above picture makes a lot of sense theoretically. Because of a clean separation between the elementary and composite sectors, it is quite easy to imagine how such a structure can be embedded into a UV complete framework (a high-scale CFT, whose global symmetry group is partly gauged by the SM gauge group, and with bilinear mixings between elementary SM fermions and CFT fermionic operators). 

I would like to comment upon the Little Higgs models. In the Higgs sector, these models are similar to the Composite Higgs since both of these scenarios are described by a non-linear sigma-model of a spontaneously broken global symmetry, with Higgs(es) as PNGB.
However, in my personal opinion, Little Higgs loses conceptual clarity of Composite Higgs in the gauge and fermion sectors. The gauge sector of the SM has to be extended, so that a bigger subgroup of the global group is gauged. Also SM top left multiplet is extended. What are these extra gauge bosons and top partners, are they elementary or composite? Since they reside in the same multiplets with SM particles, we are inclined to think of them as elementary. On the other hand, they do know something about the global symmetry of the strong Higgs sector\ldots\ In this sense, the separation line between the elementary and composite sectors is not clearly drawn. Imagining a UV-completion is then more difficult. Also, most models do not go beyond the top Yukawa coupling, and so a full discussion of flavor effects is not possible. Little Higgs has been an ambitious idea but along the way to a working implementation it has lost a lot of its initial appeal. 

Coming back to Composite Higgs models, they allow for an honest and rather complete discussion of expected flavor effects, which are typically safely below the experimental bounds except for some modest tension with $\epsilon_K$. The basic reason why the FCNC effects are small is that they are suppressed by the elementary-composite mixing angles. The limits are strongest for the light generations, and this is precisely where the effects are most suppressed. This is often called ``RS GIM mechanism'', not really a great name because it has nothing to do with the original GIM mechanism except that both suppress FCNC.

Much of the literature on Composite Higgs models in the last 10 years has been phrased in the language of warped extra dimensions. However, more recently it has become increasingly clear that it's a red herring. Things can be understood quite clearly in a standard language of four-dimensional effective field theories, matching to a general four-dimensional CFT at a high scale. Working with warped extra dimensions corresponds to a particular case when the CFT is a large $N$ theory with a dual AdS description. However, there is no reason why Nature should be large $N$ (and electroweak precision tests actually point to small $N$). One could also compare warped extra dimensions to CMSSM in the sense that both are very particular realizations of a much wider idea, and focussing on this realization results in a loss of generality. 

Now, I was told by some theorists that they keep presenting Composite Higgs models to experimentalists in the language of warped extra dimensions because they think that that's what experimentalists find cool to see. So next time you hear about warped extra dimensions, beware: you may be duped.

\begin{figure}[htbp]
\begin{center}
\raisebox{0em}{\includegraphics[scale=0.5]{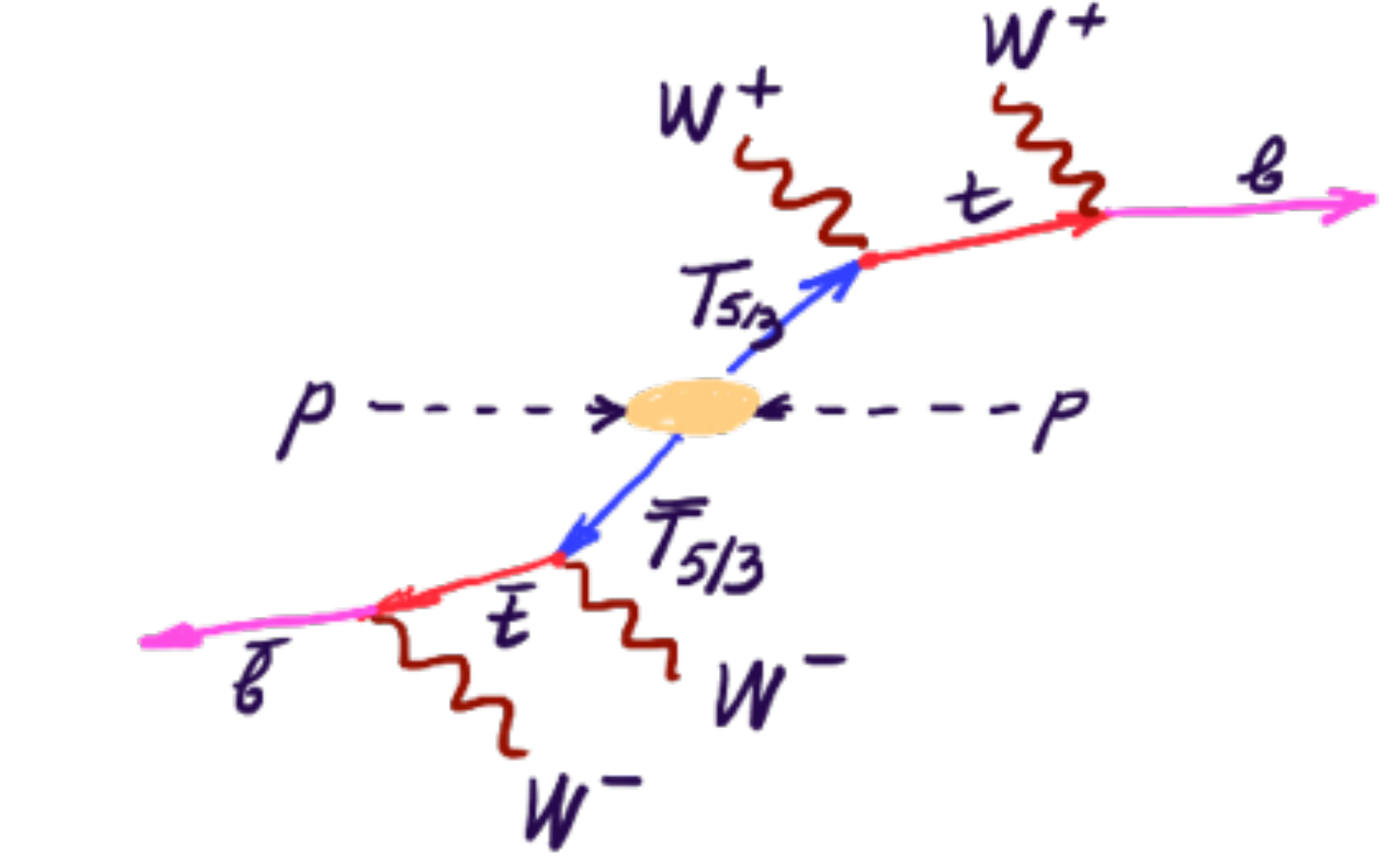}}
\caption{Exotic charge $5/3$ composite top partner quark pair production and decay. See \cite{CS} for  studies of LHC discovery potential.}
\label{T53}
\end{center}
\end{figure}

Finally, I discuss some non-Higgs collider signals of Composite Higgs. One generic feature are new quarks (composite top partners) with 500 GeV$\div$1 TeV mass and possibly exotic electric charges, like 5/3 (Fig.~\ref{T53}). From the point of view of searches, these quarks do not look very different from the 4th
generation. However, theoretically they are much less problematic: being non-chiral, they do not give problems with the electroweak precision tests. Nor do they give you a Higgs gluon fusion cross section which jumps out of the window.

\begin{figure}[htbp]
\begin{center}
\raisebox{0em}{\includegraphics[scale=0.5]{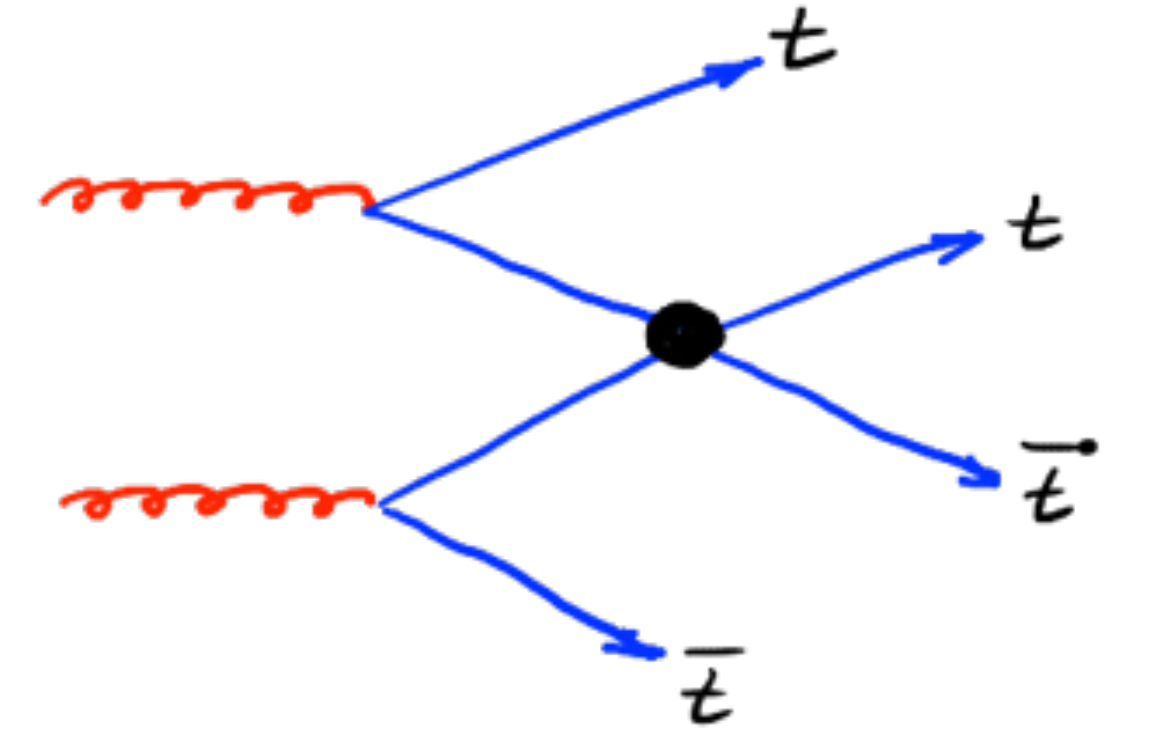}}
\caption{Anomalous $tt\bar{t}\bar{t}$ production due to top compositeness \cite{4t}.}
\label{tttt}
\end{center}
\end{figure}

Another curious signal is anomalous $tt\bar{t}\bar{t}$ production at high invariant mass due to the four-Fermi operators in the top sector (Fig.~\ref{tttt}). Notice that since only the top can be composite at a non-negligible level, existing tight bounds on light quark compositeness do not constrain this scenario.

\section{Final Remarks and Conclusions}
We have seen many impressive new physics limits set at this conference. But, have we ever truly believed in the models that are being pushed away?
$Z'$, CMSSM, split SUSY, to name a few? 

I myself certainly never believed in these. Take $Z'$. In spite of what you may have heard, this is a completely unmotivated extension of the SM. It solves nothing of its problems and has nothing to do with Naturalness. Same for split SUSY, anathema to Naturalness. CMSSM is the only victim on the list for which I feel sorry, but we can't give up on SUSY just because this straightjacketed version of it failed.

Another early casualty has been the Large Extra Dimensions scenario. But again, this was hardly a bona fide solution to the hierarchy problem. The mechanism which cuts off the Higgs mass quadratic divergence has not been concretely specified. It's only because the idea was so original that we ever gave it the benefit of the doubt. Now with LHC limits on the $(4+n)$-dimensional Planck scale already a factor two above the Tevatron limits, it's basically gone.

The truth is, apart from SUSY, there are only two other motivated scenarios for TeV-scale physics: strong EWSB and Composite Higgs. I mentioned some of the signals expected in these models. Unlike CMSSM, they typically require much higher luminosity to be seen.

What we learned at this conference is that Nature refuses to be \emph{ad hoc}. It does not go along with models specifically constructed to have an early and/or spectacular signal. The really interesting question is whether Nature cares about Naturalness. But the answer to this question is still some years away.
 
\begin{center} 
{\bf Acknowledgements} 
\end{center}
I would like to thank Riccardo Barbieri and Riccardo Rattazzi, discussions with whom over a number of years greatly influenced my point of view on the EWSB problem. My work is supported in part by the European Program ``Unification in the LHC Era", contract PITN-GA-2009-237920 (UNILHC).

\end{document}